\documentclass[a4paper,fleqn,usenatbib,useAMS]{mnras}

\usepackage{graphicx}	
\usepackage{amsmath}	
\usepackage{amssymb}	
\usepackage{multicol}        
\usepackage{bm}		
\usepackage{pdflscape}
\usepackage{mathtools}
\usepackage{array}
\usepackage[english]{babel}
\usepackage{natbib}
\usepackage{lipsum}
\usepackage{xfrac}
\usepackage{siunitx}
\DeclareSIUnit \parsec {pc}
\DeclareSIUnit \km {km}
\graphicspath{{images/}}

\usepackage[T1]{fontenc}
\usepackage{ae,aecompl}

\usepackage{times,txfonts}

\title[]{Pseudo--Evolution of Galaxies in $\Lambda \text{CDM}$ Cosmology}

\author[Vasanth Balakrishna Subramani et al.]{Vasanth Balakrishna Subramani$^1$\thanks{Contact e-mail: \href{mailto:Vasanth}{bs.vasanth@gmail.com\text{, }pkroupa@uni-bonn.de}}\thanks{Present address: Helmholtz-Institut f\"ur Strahlen- und Kernphysik (HISKP), Universit\"at Bonn, Nussallee 14-16, D-53115 Bonn, Germany},
Pavel Kroupa$^{1,3}$,
Hossein Shenavar$^2$, 
Vyoma Muralidhara$^{4}$
\\
$^{1}$Helmholtz-Institut f\"ur Strahlen- und Kernphysik (HISKP), Universit\"at Bonn, Nussallee 14-16, D-53115 Bonn, Germany\\
$^{2}$Department of Physics, Ferdowsi University of Mashhad, P.O. Box 1436, Mashhad, Iran\\
$^{3}$Charles University in Prague, Faculty of Mathematics and Physics, Astronomical Institute, Hole\v{s}ovi\v{c}k\'ach 2, 180 00 Praha 8, Czech Republic\\
$^{4}$ Argelander-Institut f\"ur Astronomie, Universit\"ut Bonn, Auf dem H\"ugel 71, 53121 Bonn, Germany}
\date{accepted July 2019}

\pubyear{2019}
\begin{document}
\label{firstpage}
\pagerange{\pageref{firstpage}--\pageref{lastpage}}
\maketitle

\begin{abstract}
Our knowledge about galaxy evolution comes from transforming observed galaxy properties at different redshifts to co-moving physical scales. This transformation depends on using a cosmological model. Here the effects of unintentional mixing of two different cosmological models on the size evolution of galaxies is studied.  As a gedanken experiment, a galaxy of fixed proper size and luminosity is moved across different redshifts. The apparent size of this galaxy is then interpreted with a cosmological model presumed by the observer, which is different compared to the cosmology exhibited by the Universe. In such a case, a spurious size evolution of the galaxy is observed. A galaxy behaving according to the $R_{\text{h}}=ct$ and Neumann's cosmology, when interpreted with the $\Lambda$CDM cosmological model, shows an increase in size by a factor of 1.1 and 1.3 from $z=7.5$ to $z\approx 0$, respectively. The apparent size of a galaxy in a static Euclidean cosmology, when interpreted in the $\Lambda$CDM model, shows a factor of 23.8 increase in size between $z=7.5$ to $z\approx0$. This is in close agreement with the observational data with a size increase of a factor of 6.8 between $z=3.2$ to $z\approx 0$. Furthermore, using the apparent size data, it is shown that the difference between the derived proper sizes in $R_{\text{h}}=ct$, Neumann's and $\Lambda$CDM  cosmological models are minimal.
 \end{abstract}

\begin{keywords}
galaxies: evolution \texttt{--} galaxies: distances and redshifts \texttt{--} cosmology: theory \texttt{--}  cosmological parameters \texttt{--} {gravitation}
\end{keywords}




\section{Introduction}
Galaxies are born and evolve subject to cosmological boundary conditions. At the same time, the physical parameters describing a galaxy at any cosmological epoch can only be determined by applying a pre-supposed cosmological model in the interpretation of the observed, apparent parameters. Our knowledge of cosmology and our knowledge of galaxy formation and evolution are thus interrelated. With this contribution we aim to elucidate this issue, raising the question how reliable our understanding of galaxy evolution is, given that some doubts on the validity of the standard $\Lambda$CDM cosmology have been raised (see below).

There are a number of geometrical tests -- Hubble diagrams \citep*{1986ApJ...301..544L,1997ApJ...477L..17S,2013AdAst2013E...8M}, Tolman surface–brightness tests \citep{2001AJ....122.1084L,2006AIPC..822....3A,2006AIPC..822...60L,2014IJMPD..2350058L}, angular
size tests \citep{Kapahi,Kellermann,Lopez_2010} -- to deduce the correct cosmological model. However, almost all these tests are affected by the unknown physical size evolution of galaxies at high redshift. The observationally deduced proper sizes of galaxies are in addition six times smaller at $z=3.2$ than at $z=0$ \citep{Lopez_2010}. Such a strong size evolution is poorly understood despite numerous attempted explanations relying on e.g. higher densities \citep*{1998MNRAS.295..319M,2004ApJ...600L.107F,2006ApJ...650...18T}, luminosity evolution \citep{2003MNRAS.343..978S}, mergers \citep{2006ApJ...650...18T, 2007MNRAS.382..109T}, or massive outflows due to quasar feedback \citep*{2008ApJ...689L.101F}. However, the static Euclidean cosmological model fits the data without any necessity of a size evolution of a galaxy \citep{2018MNRAS.477.3185L}.

In addition, the angular size test is also used to constrain the various parameters of cosmological models. \citet{2014IJMPD..2350058L} argue that the observed UV surface brightness of galaxies are inconsistent with an expanding Universe but are consistent with a static Euclidean Universe. This application of the Tolman test on the data leaves the physical origin of the observed galaxy redshift unaccounted for. \citet*{Wei} used the galaxy angular size with redshift to study $\Lambda \text{CDM}$ and $R_{\text{h}}=ct$ cosmological models. These authors have shown that the best-fitting values of the $\Lambda \text{CDM}$ model occur when $(\Omega_{m} , H_{0}) = (0.50, 73.9^{+10.6}_{-9.5} \si{\kilo\meter\per\second\per\mega\parsec})$. This estimated matter share of the Universe is significantly different from the concordance parameter $ \Omega_{ m} \approx 0.3$ derived by the \cite{2016A&A...594A..13P}. It is also shown that, although both models appear to fit the data based on their $\chi^{ 2}$, Bayes Information Criterion (as the selection tool) favours $R_{\text{h}}=ct$  over $\Lambda \text{CDM}$ with a likelihood of $86 \%$ versus $14 \%$. Also prior studies have shown that unless one presumes a surprisingly large increase  in the size of elliptical and disk galaxies (a growth factor of $\approx 6$ from $ z \approx 3$ to 0) they do not match these models. Therefore \citet{Wei} have concluded that because $\Lambda \text{CDM}$ cosmology fits the data as well as $R_{\text{h}}=ct$ model, the large growth rate might be due to  some other astrophysical phenomena such as mergers and/or selection effects rather than an incorrectly applied expansion. For example, \citet{2019arXiv190410992S} find that the half-light radius grows much more rapidly with time than the half-mass radius which shows little evolution over the redshift range $z\approx 2.5$ to 1.5.

Numerous studies have tried to implement ultra -- compact radio sources as a better standard rod. As reasoned by \citet{Kellermann} and \citet*{Gurvits_Kellermann_Frey} the underlying physics of these ultra--compact radio sources is mostly governed by a limited number of physical parameters, such as the mass of the central black hole, the accretion rate. Thus compact radio sources are less prone to evolutionary effects than extended radio sources. \citet*{Zhu_Fujimoto} used the same data as \citet{Gurvits_Kellermann_Frey} of the compact radio sources and X-ray gas mass fractions of galaxy clusters to determine the equation of state of dark energy. They have found that in a flat universe, the equation of state of dark energy is bounded as $-2.22 < w_{x} < -0.62$ at 95.4 $\%$ confidence level. This represents a nontrivial lower bound on $w_{x} $. 

\citet{Lopez_2010} and \citet{Paschenko_2011} have questioned the validity of the ultra-compact radio sources as standard rods. 
  Using the ultra-compact radio sources, \citet{Kellermann} suggests the Einstein-de Sitter model to be a more viable model, whereas \citet{1997MNRAS.285..806J}, using a much larger sample with a very different measure of angular size, and at a lower frequency, state it is not compatible with Einstein-de Sitter but with a solution $\Omega_m = 0.2$ and $\Omega_\Lambda = 0.8$. \cite{Jackson04} and \cite{JJ06} update and confirm this result using an expanded data set.  \citet{Paschenko_2011} suggest that the resolution of the ground based VLBI data and their dependence on the flux has an effect on the luminosity-linear size correlation. They conclude that it is not feasible to use ultra-compact radio sources as standard rods, a criticism challenged by \cite{Jackson12}. 

Although, baryonic acoustic oscillations in cosmic microwave background radiation or large scale structure are considered as the standard rod for cosmic expansion \citep*{2012A&A...540A.115R}, it does not have any relevance on the standard rod in the context of galaxy size evolution. Dust emission at cosmological distances may significantly affect the CMB \citep{Vavrycuk} such that the Planck solution to the $\Lambda \text{CDM}$ model may need to be revised, and significant problems of the $\Lambda \text{CDM}$ model have been documented on all scales \citep{Kroupa2010,2010Natur.465..565P,2012LRR....15...10F,Kroupa2012,Kroupa2015,2019arXiv190503258H}. The significant tension between the locally-measured-Hubble-Lemaitre constant and the value predicted using the standard $\Lambda \text{CDM}$ model and Planck satellite measurements of the CMB is becoming an unsurmountable problem \citep{2019arXiv190307603R}. Thus the $\Lambda \text{CDM}$ model may not be the correct description of the Universe \citep{Kroupa2012,2014IJMPD..2350058L, 2018MNRAS.477.3185L, 2018A&A...610A..87M}.   

The current circumstances thus make it conducive to perform a gedanken experiment by taking a hypothetical standard rod of, for example, 50 kpc in order to study the dependence of the inferred apparent size on the cosmological model. For this a theoretical galaxy of radius 25 kpc is moved through different redshifts and its apparent size is interpreted via $R_\text{h}=ct$, Neumann and $\Lambda \text{CDM}$ cosmological models. This enables to quantify the actual dependence of the apparent size on the cosmological model independent of any size evolution mechanism. As an example, using the data of apparent size and redshift from table 1 of \citet{2018MNRAS.477.3185L}, the cosmology dependence on interpreting the proper size and  luminosity from their apparent measurements is illustrated here. Furthermore, the effect of using a particular pre-supposed cosmological model to study the apparent size of a galaxy that is in reality immersed in another cosmological model is demonstrated. A spurious size evolution of galaxies is observed.

The outline of this paper is as follows: Section \ref{section3} develops the Hubble-Lemaitre parameter in different cosmological models. Section \ref{section4} compares the angular diameter distance and luminosity distance in different cosmologies and subsequently its effect on the deduced apparent radius and apparent magnitude of a theoretical galaxy. Using data from table 1 of \citet{2018MNRAS.477.3185L} the dependence of cosmology in deducing the proper size of galaxies is exemplified in Section \ref{Data_analysis}. In Section \ref{pse} and Section \ref{non-exapnding-model}, the pseudo\texttt{--}evolution of galaxies with redshift  in an expanding and non-expanding universe is introduced, respectively. Section \ref{Conclusion} concludes with results and future prospects. The Appendix gives a brief introduction to cosmological perturbation theory used to arrive at Neumann's cosmological model.

\section{Hubble's parameter}  \label{section3}

The Hubble-Lemaitre (H-L) parameter, H, forms the crux of any expanding model of the Universe. It is essential to arrive at the notion of distance, and further to derive apparent magnitude and apparent size. Hence, in this section a brief introduction to the H-L parameter is given for Neumann, $R_{\text{h}}=ct$ and $\Lambda \text{CDM}$ cosmological models because this is not available in the current literature.   
The Neumann cosmological model is based on using the Neumann boundary condition on Eq. \ref{laplace}  derived from cosmological perturbation theory in the Appendix \citep{2016Ap&SS.361...93S}. In Neumann's model, the Friedmann metric is used as the background metric along with a scalar perturbation. In order to arrive at the H-L parameter in Neumann's model it is necessary to first determine the ${^l}G_{00}$ component of Einstein's field equation for the background metric, which is given in Eq. \ref{l^G_00} in the Appendix. Taking the ${^l}G_{00}$ component in Eq. \ref{l^G_00} and substituting for $\Phi$ using Eq. \ref{solution_to_laplace} and collecting the terms proportional to $c_1$ to the right hand side of terms obtains
\begin{equation}
\frac{12 \Psi H^2}{R^4}+\frac{2 \bigtriangledown^2 \Psi }{R^4}= 4\pi R^2 G \delta T^{0}_{0}+\frac{6 c_1 H^2}{R^2}.
\end{equation}
Now, taking the terms proportional to $c_1$ and adding them to the $ {^b}G_{00}$ component of Einstien's field Eq. \ref{backgroundEinsteintensor}, the non-standard Friedmann's equation results, 
\begin{equation}
    3H^2= 8 \pi G \rho + \frac{6 c_1 H^2}{R^2},
    \label{modified_friedmann}
\end{equation}
along with the constant $c_1$ term contributing to the energy density. Here the additional term with $c_1$ in Eq. \ref{modified_friedmann} is rearranged to get $8 \pi G \rho_{c_1}$, where $\rho_{c_1}$ referred to as the Neumann term. The $\rho_{c_1}$ also contributes to the overall energy density along with the conventional energy density $\rho$, where $\rho$ is composed of baryonic matter density, $\rho_m$, and radiation density, $\rho_r$. The interpretation for the constant $c_1$ is still an open question. For an experimental determination of the value of $c_1$ using strong lensing systems see \citet{2016Ap&SS.361...93S}. This derivation is different from eq. 57 in the appendix  of \citet{2016Ap&SS.361...93S} since the correct expression of ${^l}G_{00}$ from eq. 7.38 in \citet{Mukhanov} is used, and $\Phi= \Psi -c1$ is substituted instead of $\Psi$ as given in eq. 20 of \citet{2016Ap&SS.361...93S}. Therefore, this derivation also serves as an erratum for the derivation of eq. 57 of \citet{2016Ap&SS.361...93S}. 

The energy densities at time $t$ can be written into the more favoured dimensionless density parameters 
\begin{align}
  \Omega_i & \equiv \frac{8 \pi G \rho_i(t)}{3 H^2(t)}, 
 \end{align}
 and
\begin{align}
 \Omega_k & \equiv -\frac{k}{H^2(t)R^2(t)},
 \end{align}
 where k= -1, 0 and 1 for a hyperbolic, flat and closed universe, respectively.

Writing the Friedmann Eq. \ref{modified_friedmann} in terms of $\Omega_{0,i}$, 
   \begin{align}
        H^2(t)= H_0 \bigg( \frac{\Omega_{0,m}}{a^3}+\frac{\Omega_{0,r}}{a^4}+\frac{\Omega_{0,k}}{a^2}+\Omega_{0,c_1}\frac{H}{H_0} \bigg),
        \label{HUbble}
    \end{align}
in which $\Omega_{0,i}$ is the i$^{th}$ density parameter at the present time $t_0$, while
\begin{equation}
a(t)\equiv\frac{R(t)}{R_0}=(1+z)^{-1},
\end{equation} 
is the normalized scale factor.
Here $\Omega_{0,m}$, $\Omega_{0,r}$, $\Omega_{0,k}$ and $\Omega_{0,c_1}$ are the baryonic matter, radiation, curvature and Neumann densities of the universe, respectively. The curvature density parameter is set at zero assuming the universe to be flat, and the value of the Neumann term is obtained using the flatness condition $\Omega_{tot}=\sum_i\Omega_i (t)=1$. Here, $\Omega_{0,m}$ represents only the baryonic matter, not inclusive of dark matter, unlike in the $\Lambda \text{CDM}$ cosmological model, to give the Neumann cosmological model parametric freedom. By solving the quadratic Eq. \ref{HUbble} and taking only the positive solution \footnote{The negative sign corresponds to a contracting universe.} for the square root we get the Neumann H-L parameter,
\begin{equation}
    H_{Neu}(t)= \frac{H_0}{2} \bigg( \Omega_{0,c_1} + \sqrt{\Omega^2_{0,c_1}+4(\frac{\Omega_{0,k}}{a^2}+\frac{\Omega_{0,m}}{a^3}+\frac{\Omega_{0,r}}{a^4})} \bigg).
\label{Neuhubble}
\end{equation}

 In the $R_{\text{h}}=ct$ cosmological model the universe expands with the constant velocity of speed of light $c$ \citep{Melia_Fulvio_Shevchuk}. Therefore, in this case the scale factor is $a=ct$, and the H-L parameter,
 \begin{equation*}
 H=\frac{\dot{a}}{a},
 \end{equation*}
 becomes $H=\sfrac{1}{t}$. Substituting for $t$ via the expression $1+z=\frac{a(t_0)}{a(t)}=\frac{1}{t}$, where $t$ and $t_0$ are emission and observation time, respectively, gives the H-L parameter in the $R_{\text{h}}=ct$ cosmological model as \citep{2017ApJ...835..270W,1475-7516-2017-11-029}
 \begin{equation}
     H_{R_h}= H_0 (1+z).
 \end{equation}
 Here $H_0$ is the current value of the H-L parameter, and it is also the only sole free parameter.
 
\begin{figure}
\includegraphics[width=9cm]{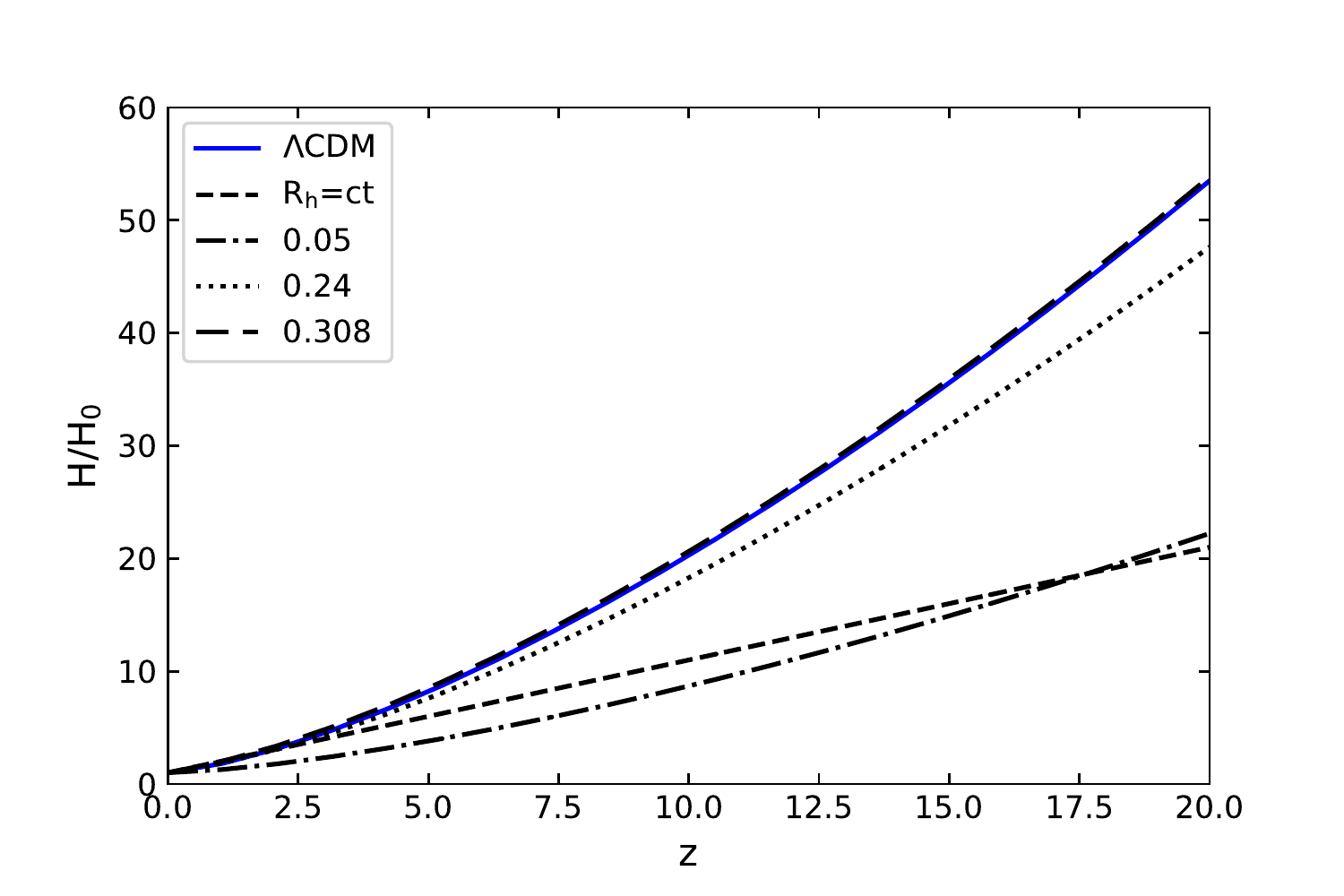}
\centering
\caption{The blue curve shows the H-L parameter with respect to redshift for the $\Lambda \text{CDM}$ cosmology. The Neumann cosmology is shown with different baryonic matter densities and the H-L parameter for the $R_{h}=\text{ct}$ model is the linear relation shown as a dashed straight line. This figure is normalized to the current value of the H-L parameter as obtained by the recent \citet{2016A&A...594A..13P}.} 
\label{hubble constant}
\end{figure}

The H-L parameter in the $\Lambda \text{CDM}$ model is given by  
\begin{equation}
    H^2_{\Lambda CDM}(t)= H^2_0 \, \bigg(\frac{\Omega_{0,k}}{a^2}+\frac{\Omega_{0,m}}{a^3}+\frac{\Omega_{0,r}}{a^4}+\Omega_{0, \Lambda} \bigg),
\end{equation}
where $\Omega_{0,m}$, $\Omega_{0,r}$, $\Omega_{0,k}$, and $\Omega_{0,\Lambda}\equiv \frac{\Lambda}{3 H^2_0}$ are the present-day matter (including the dark matter), radiation, curvature and dark energy density parameters, respectively, with $\Lambda$ being a cosmological constant associated with dark energy. The value of $H_{+ \Lambda \text{CDM}}$ or $H_{- \Lambda \text{CDM}}$ gives an expanding or contracting universe, respectively. Throughout this paper $H_{+}$ and a flat universe is assumed for the Neumann, $R_{\text{h}}=ct$ and $\Lambda \text{CDM}$ cosmological models \citep{2016A&A...594A..13P}. In Fig \ref{hubble constant} the evolution of H-L parameter with respect to redshift $z$ is compared for all the models. The Neumann model is also plotted for different baryonic matter densities, one notices that for both the Neumann and $\Lambda \text{CDM}$ models, the H-L parameter coincides for $\Omega_m=0.308$, $\Omega_r=5\times 10^{-5}$ and $\Omega_{\Lambda}=0.691$, which are the values from the recent Plank
analysis \citep{2016A&A...594A..13P}. In Neumann's model $\Omega_m$ is interpreted as only the baryonic density, whereas in the $\Lambda \text{CDM}$ it is total matter density including dark matter.   

It also noted that the H-L parameter for the $R_{\text{h}}=ct$ cosmology follows a straight line as it does not depend on any components of the cosmological fluid. Although it looks similar to Milne cosmology it is distinct from it by having a non-zero energy density. 

\section{A potpourri of Cosmologies} \label{section4}
Any object at redshift $z$ and with a physical co-moving (proper) diameter $D$ [kpc] and absolute magnitude $M$ will have a diameter in radians on the sky of 
\begin{equation}
\delta \theta=\frac{D}{D_{A}(z)}
\label{apparent_size}
\end{equation}
and an apparent magnitude of 
\begin{equation}
    m = M + 5\text{ log}_{10} \frac{D_{L}}{10}.
    \label{apparent_magnitude}
\end{equation}
The deduction of the proper diameter and the absolute magnitude of the object in the sky by using Eq.\ref{apparent_size} and Eq. \ref{apparent_magnitude} depends on the cosmological model through the angular diameter distance $D_{A}$ and luminosity distance $D_{L}$, respectively. 

The general definition of the luminosity distance and the angular distance in any cosmological model is 
\begin{equation}
D_L=c(1+z)\int_{0}^{z} \frac{dz^{'}}{H(z^{'})} \text{ Mpc,}\label{Luminosity distance}
\end{equation}
and 
\begin{equation}
D_A=\frac{c}{1+z}\int_{0}^{z} \frac{dz^{'}}{H(z^{'})} \text{ Mpc,} \label{Angular distance}
\end{equation}
respectively. Therefore the distances in  $\Lambda \text{CDM}$\footnote{From here on we refer to the $\Lambda \text{CDM}$ model "as the standard cosmological model with the \cite{2016A&A...594A..13P} data."}, $R_{\text{h}}=ct$ and Neumann's cosmological models are obtained by inserting their respective definitions of the H-L parameter. As a representative example, for the $R_{\text{h}}=ct$ cosmological model \citep{Wei}  
\begin{equation} \label{HRCT}
D_{A} = \frac{c}{H_{0}}\frac{\ln{(1+z)}}{1+z},   
\end{equation}
by inserting $H_{R_h}$ in Eq. \ref{Angular distance} with $H_0$ being the current value of the H-L parameter.

\begin{figure}
\includegraphics[width=9cm]{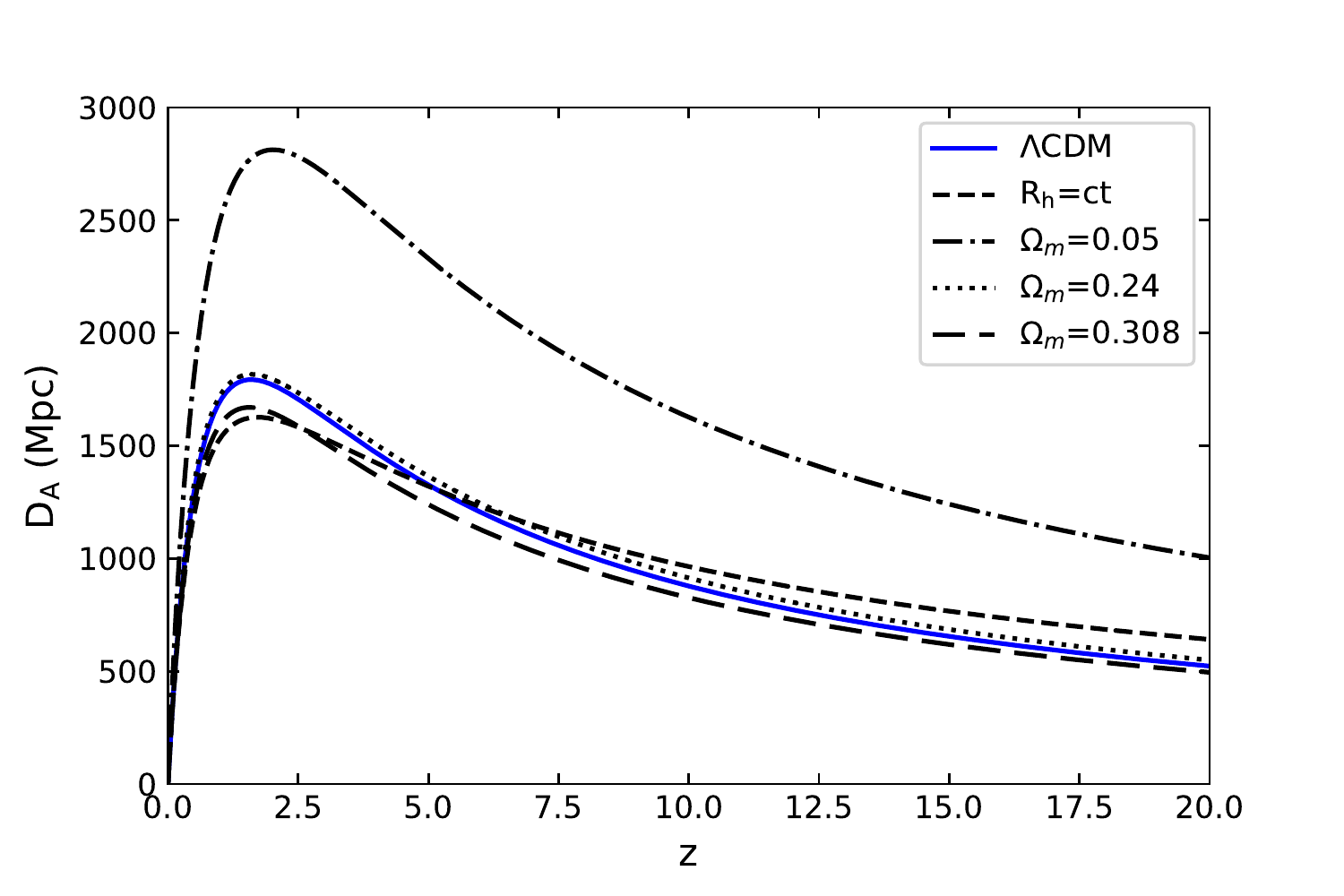}
\caption{The angular diameter distance with respect to redshift for $\Lambda \text{CDM}$, $R_{\text{h}}=\text{ct}$ and Neumann's cosmology. Symbols are as in Fig. \ref{hubble constant}. This figure also serves as an erratum for the fig. 8 in \citet{2018arXiv181005001S}.} 
\label{DA}
\centering
\end{figure}

\begin{figure}
\includegraphics[width=9cm]{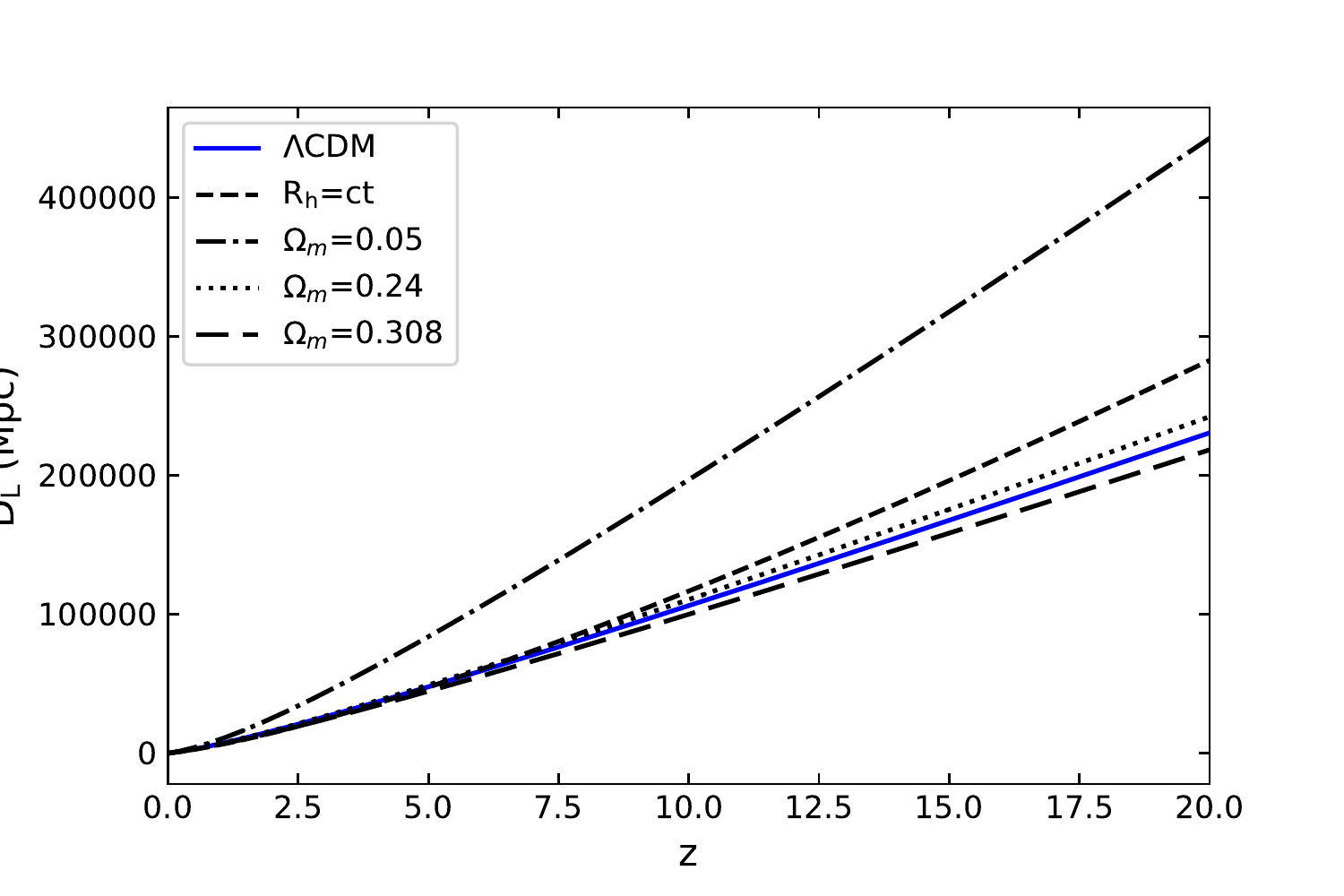}
\centering
\caption{The luminosity distance with respect to redshift for $\Lambda \text{CDM}$, $R_{\text{h}}=\text{ct}$ and Neumann's cosmology. This figure also serves as an erratum for the fig. 7 in \citet{2018arXiv181005001S}. }
\label{luminosity distance fig}
\end{figure}

\begin{figure}
\includegraphics[width=9cm]{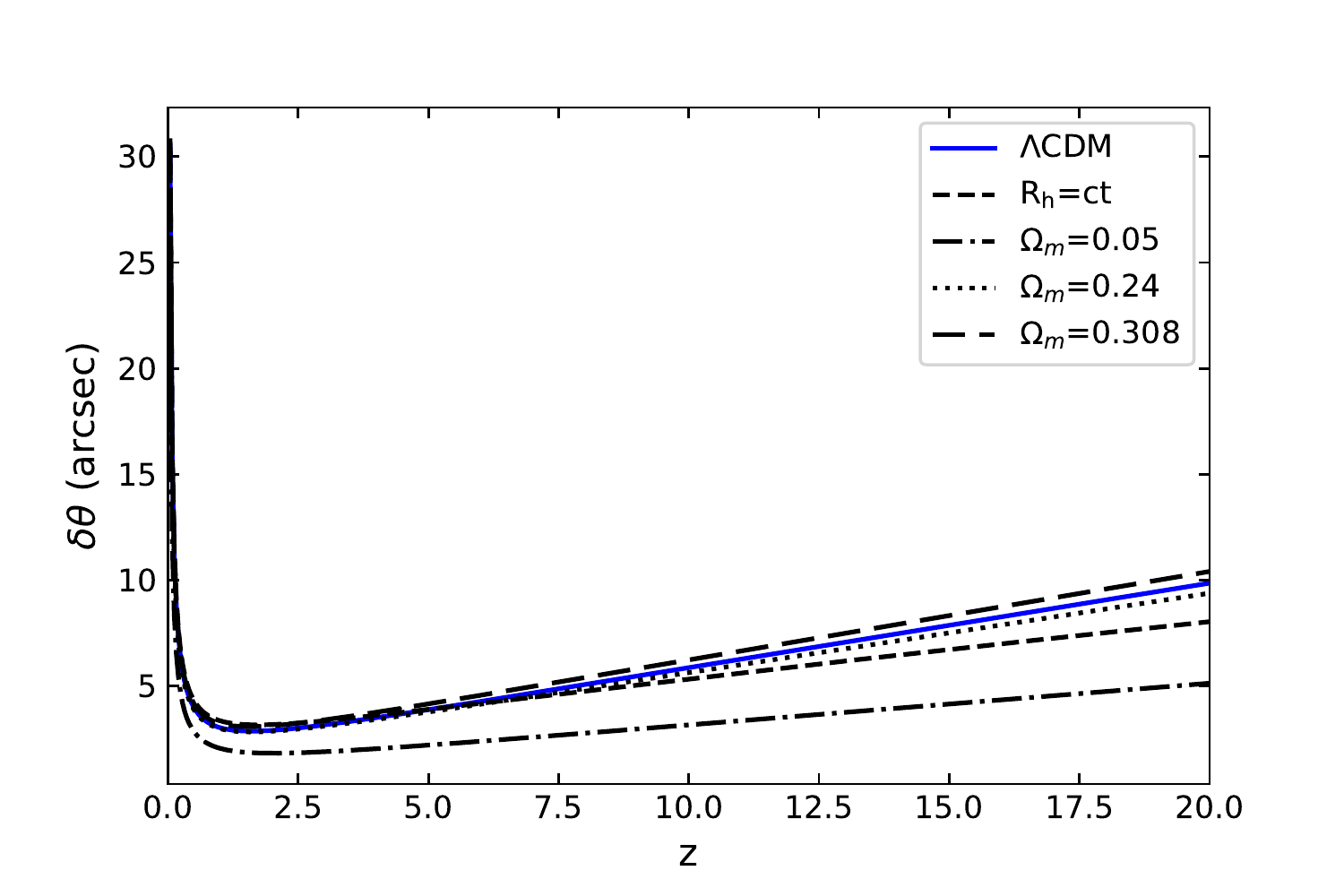}
\centering
\caption{The apparent radius of a galaxy with a proper radius of $25 \text{\, kpc}$ in the $\Lambda \text{CDM}$, the $R_{\text{h}}=\text{ct}$ and the Neumann cosmological models.}
\label{apparentsize} 
\end{figure}

Figures \ref{DA} and \ref{luminosity distance fig} show the angular diameter distance and luminosity distance for $\Lambda \text{CDM}$, $R_{\text{h}}=ct$, and Neumann's cosmological models for different baryonic mass densities. As evident in Fig. \ref{DA}, all the three cosmological models show a cosmological lensing effect, that is, the apparent size of an object might seem to increase for distances approximately farther than $z=1$ (see Fig. \ref{apparentsize}). This is the consequence of expanding universe models. Furthermore, Fig. \ref{DA} shows that for a given redshift the same object is closer in Neumann's model with baryonic matter density $\Omega_m=0.308$ compared to the $\Lambda \text{CDM}$ model, which is also modelled for a total matter density (baryonic matter and dark matter) of $\Omega_m=0.308$. For $R_\text{h}=ct$ and $\Lambda \text{CDM}$ cosmological models the distance to the same object differ only in the peak and for later redshifts. On the other hand, the distances in the Neumann model increases with decreasing baryonic matter density.   

In Figure \ref{luminosity distance fig}, the luminosity distance for the same object at a given redshift is inferred to be larger in $R_{\text{h}}=ct$ compared to $\Lambda \text{CDM}$ and Neumann's cosmological model with $\Omega_m > 0.2$. The luminosity distance for both the $\Lambda \text{CDM}$ and $R_{\text{h}}=\text{ct}$ models are distinguishable only at higher redshifts beyond $z=5$. For the Neumann cosmological model the inferred luminosity distance of an object looks closer for a universe with higher baryonic matter density. The curve with same baryonic matter density as the Planck solution seems to be nearer in the Neumann model than in $R_\text{h}=ct$ and $\Lambda \text{CDM}$ cosmological models. The consequence of these differences in the angular diameter distance and luminosity distance can lead to deducing a different proper size and absolute magnitude using the observed apparent size and apparent magnitude in different cosmological models.    

 This problem is illustrated by taking a galaxy of fixed proper radius of 25 kpc and fixed absolute magnitude $M_V=-21.5$ (apparent magnitude of the Andromeda galaxy) in the photometric V-band  \citep{2005ApJ...635L..37R}, and moving it through different redshifts, while observing the differences in the apparent measurements of the galaxies by different cosmological models.

In Figure \ref{apparentsize}, the deduced apparent radius in $\Lambda \text{CDM}$ and the $R_{\text{h}}=\text{ct}$ models are smaller than in Neumann's cosmological model with $\Omega_m=0.308$ for a given redshift of a galaxy with 25 kpc radius. The apparent size also decreases with decreasing baryonic matter densities for the Neumann cosmological model. It is also pertinent to note that if the universe behaved with a different cosmological model, suppose with the Neumann cosmological model with $\Omega_m=0.05$, then a galaxy at $z=10$ will appear to be bigger from the perspective of $\Lambda \text{CDM}$ and $R_{\text{h}}=\text{ct}$ cosmological models.

\begin{figure}
\includegraphics[width=9cm]{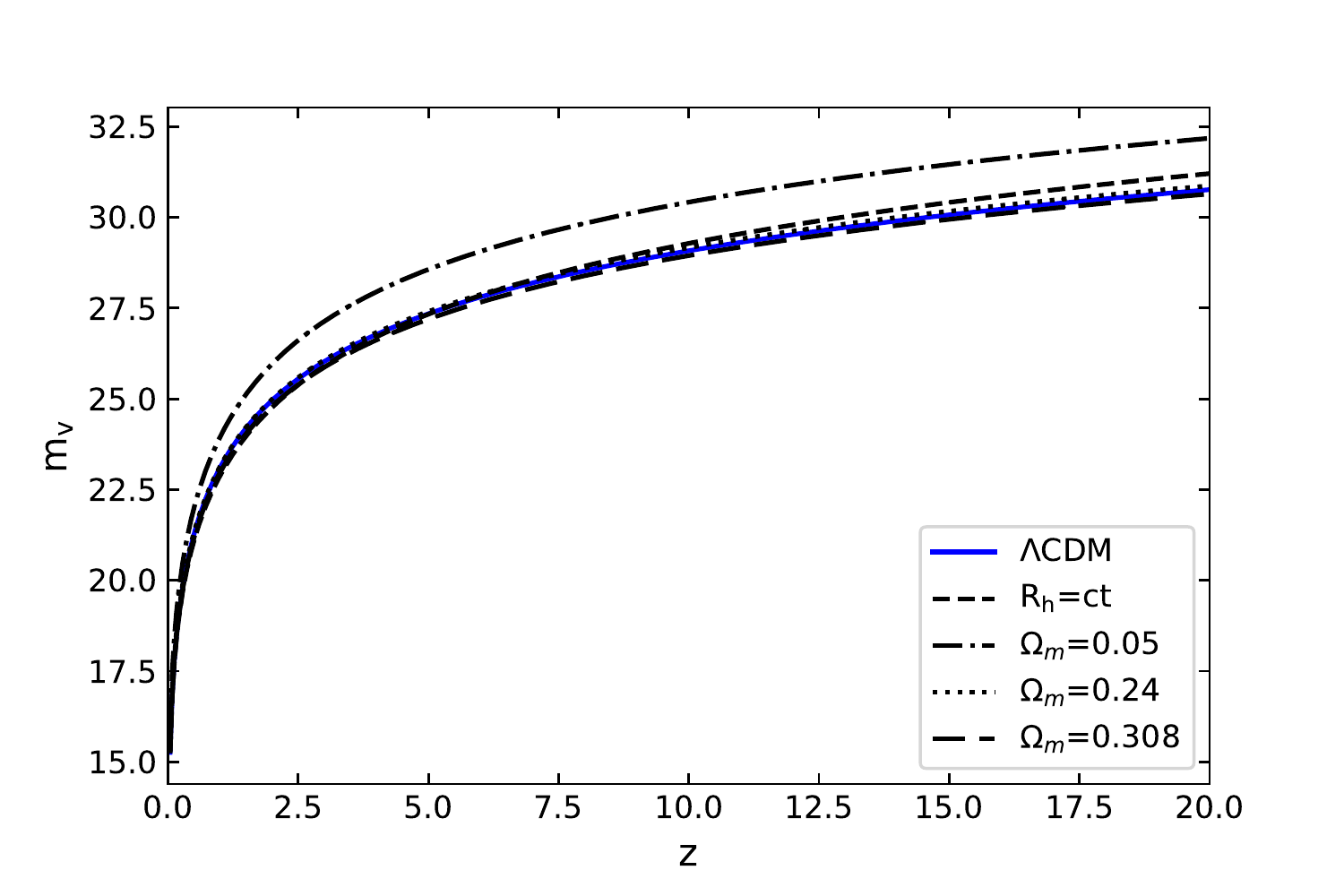}
\centering
\caption{The apparent magnitude of the Andromeda galaxy with absolute V-band magnitude $M=-21.5$ in all the three cosmological models.}
\label{apparent magnitude}
\end{figure}
 
Figure \ref{apparent magnitude} depicts the apparent magnitude of the Andromeda galaxy as interpreted by the different cosmological models. The Andromeda galaxy at very high redshift appears to be dimmer (the value of $m$ increasing) in $\Lambda \text{CDM}$ and $R_{\text{h}}=\text{ct}$ models as compared to Neumann's cosmological model with $\Omega_m=0.308$. It also appears dimmer with decreasing baryonic matter density in Neumann's cosmological model.

\section{An Example}  \label{Data_analysis}
In this section, the cosmological dependence on interpreting the proper size is studied by taking the apparent radius\footnote{The apparent radius is deduced from the proper radius in column two of table 1 from \citet{2018MNRAS.477.3185L} using the $\Lambda \text{CDM}$ cosmological model.} of the galaxies given in table 1. of \citet{2018MNRAS.477.3185L}.

\begin{figure}
\includegraphics[width=9cm]{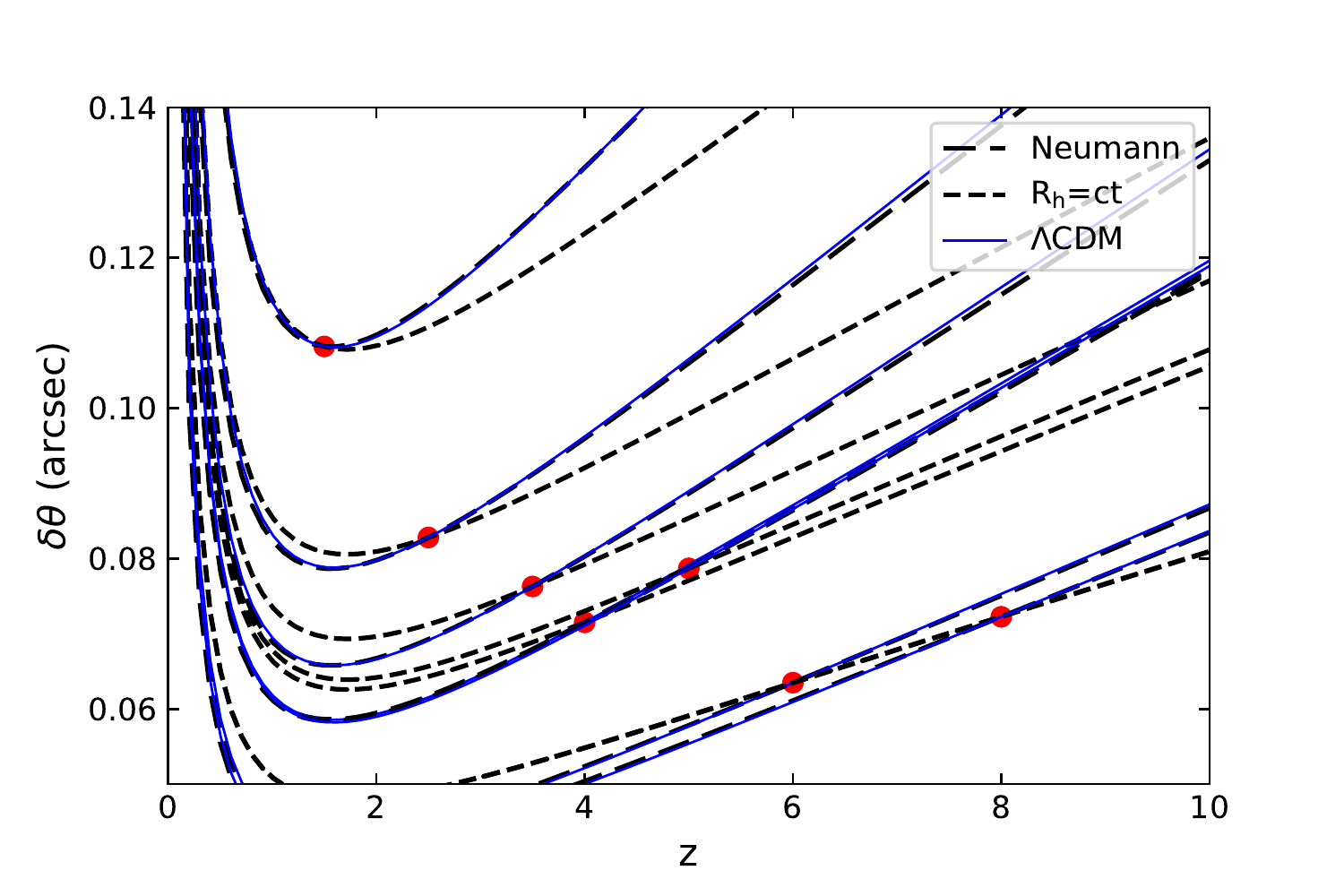}
\centering
\caption{The curves through each data point \citep{2018MNRAS.477.3185L} depict the deduced proper radius (see also Fig. \ref{Proper_size_in_all_models}) of the apparent radius of galaxies by using $\Lambda \text{CDM}$, $R_\text{h}=ct$ and Neumann ($\Omega_m= 0.308$) cosmological models. The proper sizes are derived by using the relationship between the apparent size, proper size and distance from the respective cosmological model given by 
Eq. \ref{apparent_size}. The different curves show that for the same data point the deduced proper size depends on the cosmological model the observer chooses to use.} 
\label{Proper_size_curves}
\end{figure}

In Figure \ref{Proper_size_curves}, the apparent sizes of the galaxies are fitted with their respective deduced proper sizes from the $\Lambda \text{CDM}$, $R_{\text{h}}=\text{ct}$ and Neumann's cosmological model. Figure \ref{Proper_size_curves} shows the deduced proper size curves for each data point in different models. 

\begin{figure}
\includegraphics[width=9cm]{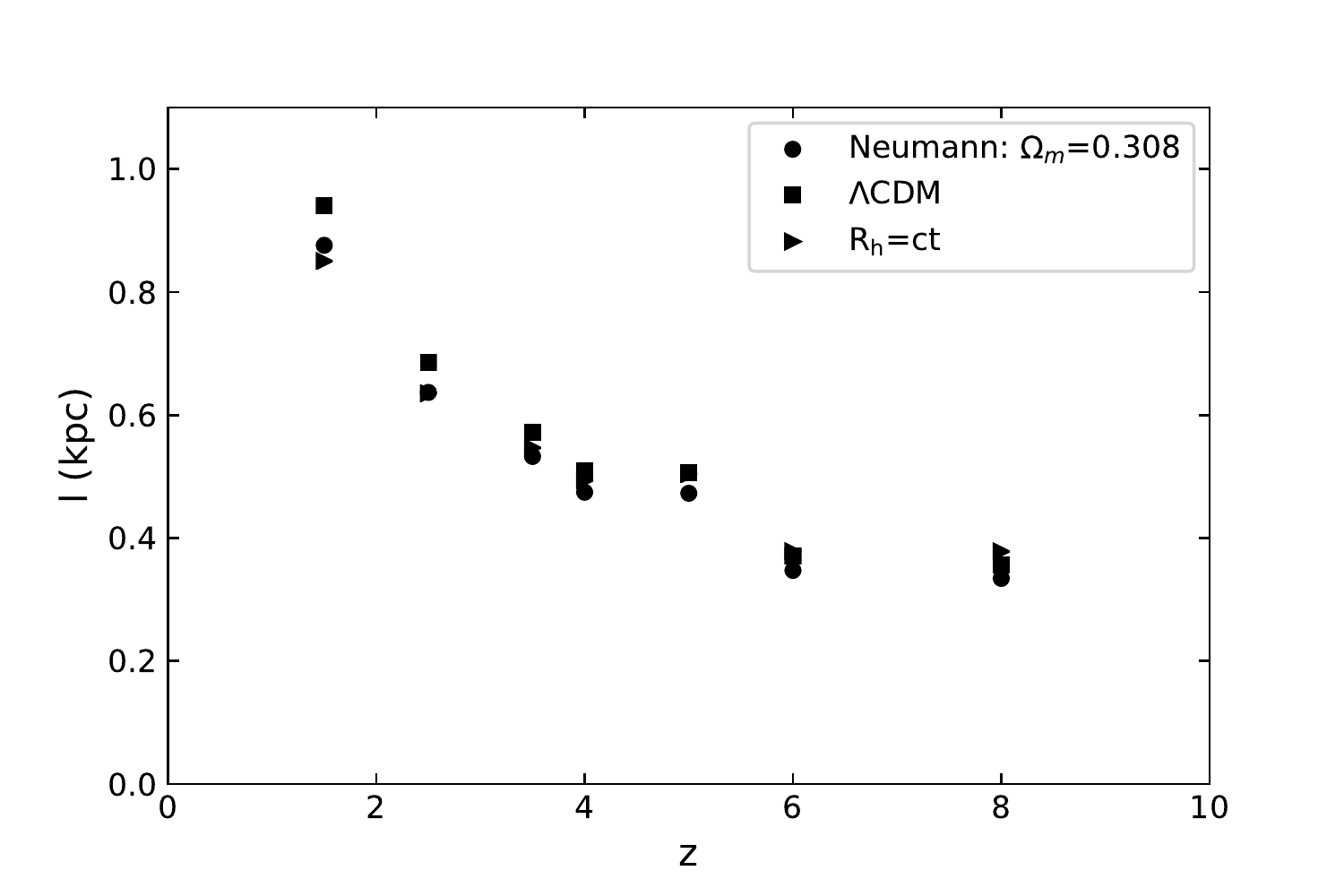}
\centering
\caption{The deduced proper radius from the apparent radius of the galaxies shown in Fig. \ref{Proper_size_curves}. Here all the cosmologies show a size evolution from the past to present, however the galaxies inflate only approximately by a factor of two between $z=8$ and $z=0$, which is significantly smaller compared to the factor of 6 between $z=3$ to $z=0$ reported by 
\citet{2004ApJ...600L.107F,Bouwens_2004,Lopez_2010,2004ApJ...604..521T,2008ApJ...676..781W,2009A&A...499...69N}. Note that the differences in the deduced proper radius in the different expanding cosmological models are minimal.}
\label{Proper_size_in_all_models}
\end{figure}

Figure \ref{Proper_size_in_all_models} depicts the deduced proper size for individual data points by the different cosmological models. In the lower redshift region upto $z=4$ the $\Lambda \text{CDM}$ and Neumann models appear to imply a larger proper size, but the proper size deduced from the $R_{h}=ct$ cosmological model dominates at $z \geq 6$. Overall the galaxies have increased approximately twice in size from $z=8$ to $z=1$. This is a small size evolution in comparison to \citet{2006ApJ...650...18T}, which concludes a size evolution by a factor of $3$ from $z=2.5$ to $z=0$. For heavier galaxies \citet{2007MNRAS.382..109T} conclude an evolution of upto a factor of $4$ from $z=1.5$ to $z=0$ and a factor of $5.5$ from $z=2.3$ to $z=0$. Other studies \citet{2004ApJ...600L.107F,Bouwens_2004,Lopez_2010,2004ApJ...604..521T,2008ApJ...676..781W,2009A&A...499...69N} suggest a size increase by a factor of approximately 6 from $z=3$ to $z=0$. See also \citet{2019arXiv190410992S} for further discussion of possible biases.

\section{Pseudo-Evolution in an Expanding Universe} \label{pse}
In Sec. \ref{Data_analysis}, the effect on the size evolution dependent only on the cosmological model is obtained in Fig. \ref{Proper_size_in_all_models}. Here the question raised is if there can be a spurious size evolution if the universe does not follow the $\Lambda \text{CDM}$ model? Is an incorrect size evolution deduced due to using an incorrect cosmological model? A possibility of such a fake phenomenon is shown by taking a universe that in reality behaves according to a Neumann or $R_\text{h}=ct$ cosmological model, and interpreting the fixed proper size of a galaxy in such a universe with the $\Lambda \text{CDM}$ cosmological model. This is achieved by theoretically generating the apparent size for a galaxy of true proper size of radius 25 $\text{kpc}$ in the Neumann and $R_\text{h}=ct$ cosmological model and deducing its proper size using the $\Lambda \text{CDM}$ model.

The relationship between the deduced proper size and theoretically generated apparent size is given by
\begin{equation}
\label{Pseudo-size-evolution}
 l_{pse} = \delta \theta_{Neu  \text{ or }  R_h} \times D_{A_{\Lambda \text{CDM}}}. 
\end{equation}
Here $\delta \theta_{Neu  \text{ or } R_h}$ is the theoretically generated apparent diameter for a galaxy of (true) proper size of radius 25 $\text{kpc}$ in Neumann or $R_\text{h}=ct$ cosmological model. The $l_{pse}$ is the proper size interpreted for a galaxy in a Neumann or  $R_\text{h}=ct$ universe via the angular diameter distance $D_{A_{\Lambda \text{CDM}}}$ of the $\Lambda \text{CDM}$ cosmological model. The apparent size and redshift are the only observed quantity for an universe behaving with a specific cosmology. The calculation here by no means advocates the correctness of any cosmological model. The motivation is rather to perform a gedanken experiment to see the consequences of using two models--one exhibited in reality by the Universe and the other used to interpret this Universe--on the size evolution of a galaxy. 

\begin{figure}
\includegraphics[width=9cm]{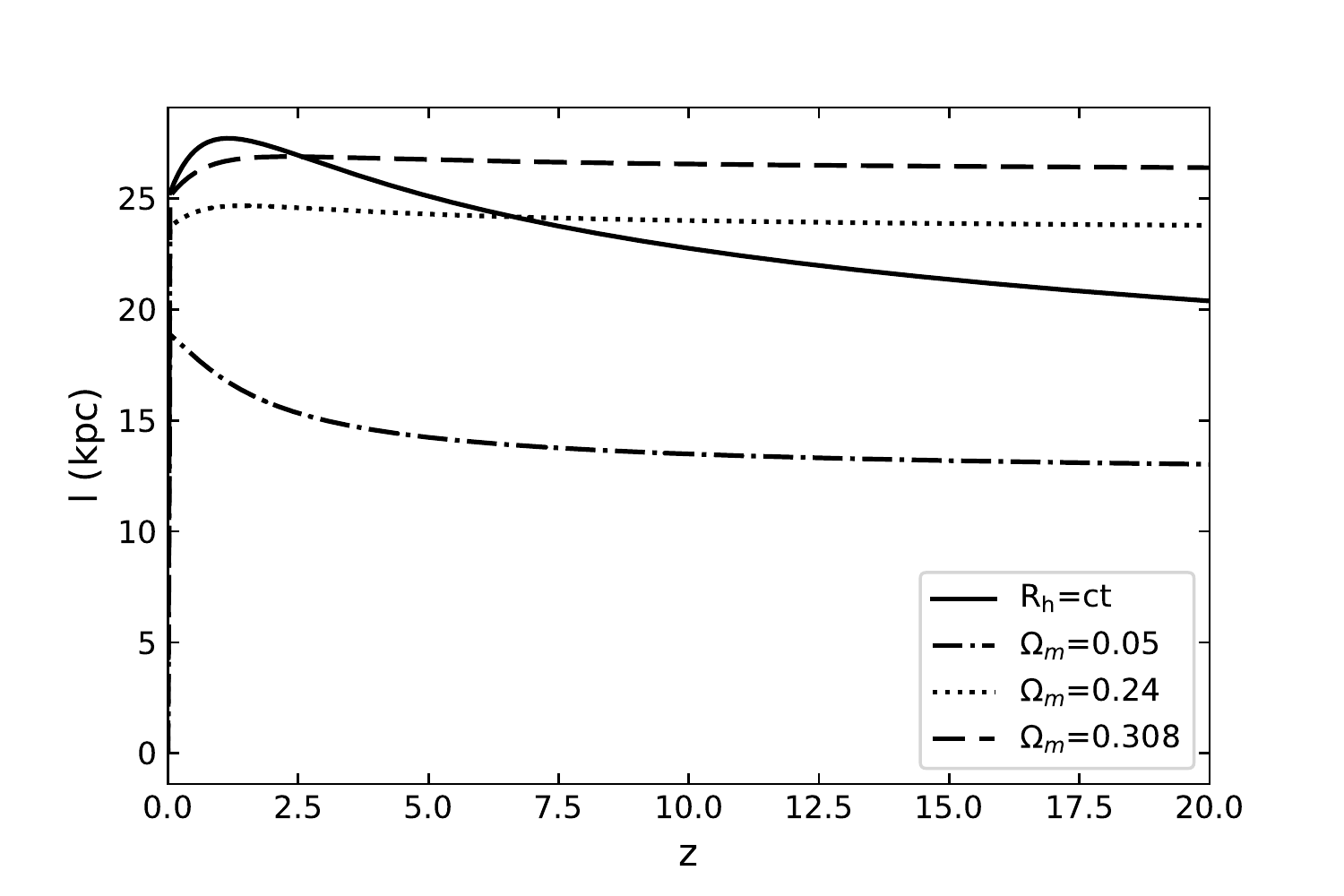}
\centering
\caption{The proper size of a theoretical galaxy of fixed radius $25 \, \text{kpc}$ with redshift. The apparent size in the assumed to be correct Neumann and $R_{\text{h}}=\text{ct}$ universe is interpreted in $\Lambda \text{CDM}$ cosmology. }
\label{size evolution}
\end{figure}

This deduced proper size $l_{pse}$ using the $\Lambda \text{CDM}$ cosmological model is shown in Figure \ref{size evolution}. Here the proper size appears to evolve differently for the galaxy in $R_\text{h}=ct$ cosmology with $H_0$ as the sole free parameter, and for Neumann's universe with baryonic matter density of $\Omega_m=0.05$, which is close to the estimated baryonic matter density of Planck solution. This size evolution is unphysical as the galaxy radius is fixed to 25 kpc in both cosmologies. Such a spurious evolution is referred to here as pseudo-evolution. Assuming the Universe is correctly described by the $R_\text{h}=ct$ model, a $\Lambda \text{CDM}$ observer would deduce the galaxy's radius to evolve from 23.7 kpc at $z=7.5$ to 26.3 kpc at $z\approx0$. This is a size evolution by a factor of 1.1 that is purely due to pseudo-evolution. A galaxy in Neumann's cosmology with $\Omega_m=0.05$ has increased in size from 13.7 kpc at $z=7.5$ to 18.4 kpc at $z\approx0$, which is an increase in size by a factor of 1.3 due to pseudo-evolution.

\section{Pseudo-Evolution in a Non-Expanding Universe}\label{non-exapnding-model}
\begin{figure}
    \centering
    \includegraphics[width=9cm]{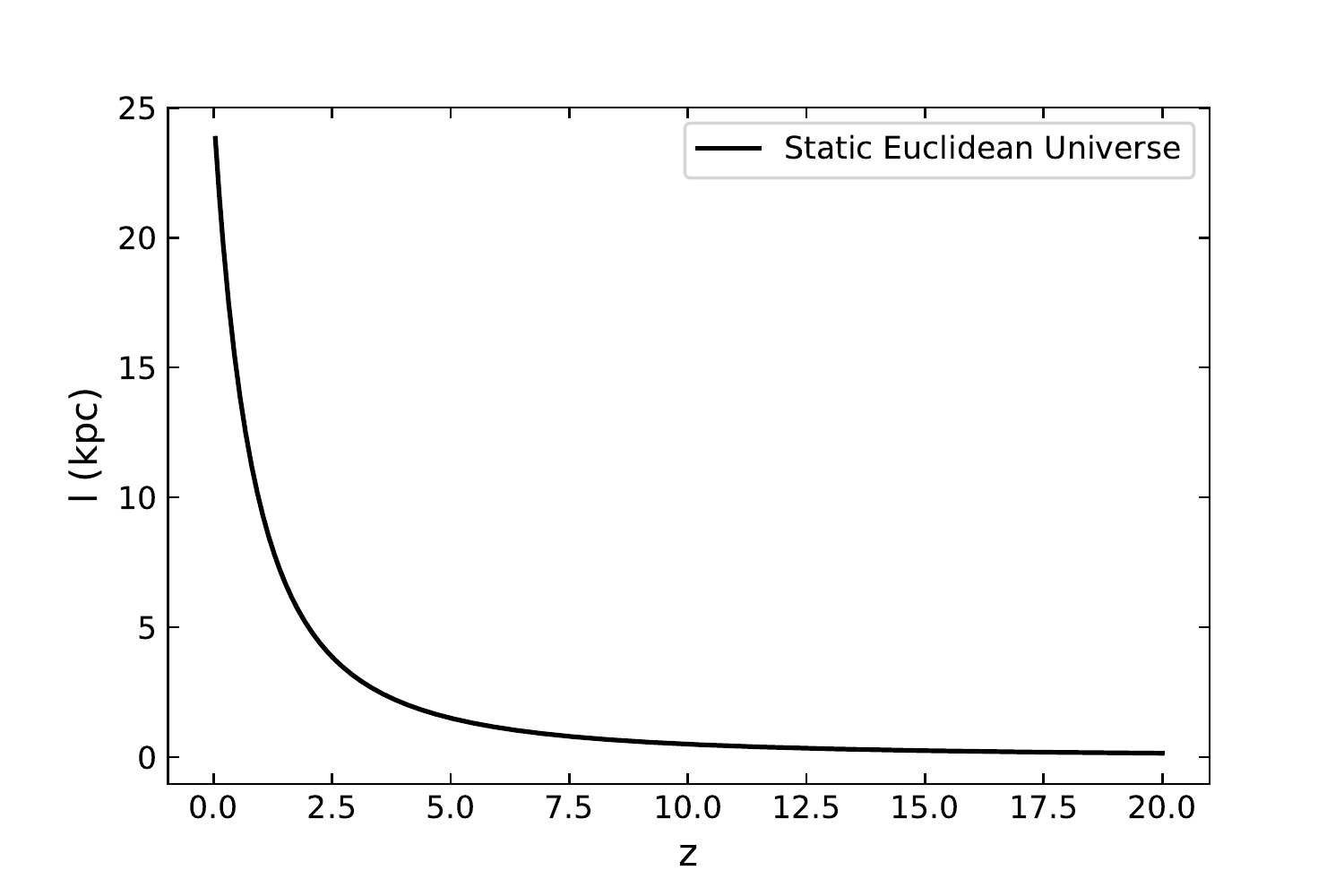}
    \caption{The pseudo-evolution of a galaxy of fixed radius, 25 kpc, in static Euclidean model when interpreted via $\Lambda$CDM cosmological model.}
    \label{Pseudo_Static}
\end{figure}

In this section, a galaxy in a non-expanding universe is interpreted in the $\Lambda$CDM cosmological model. The distance-redshift relationship for a static universe is \citep{2014IJMPD..2350058L} 
\begin{equation}
    D_A(z)= \frac{c z}{H_0}.
\end{equation}
Note that in this model the redshift is a physical phenomenon that is not related to velocity. Now, taking a galaxy of radius $25 \text{ kpc}$ in the static universe, whose apparent size is given by Eq. \ref{apparent_size}, one can arrive at the pseudo-evolution from Eq. \ref{Pseudo-size-evolution}, for interpretation via the $\Lambda \text{CDM}$ cosmological model. This is depicted in Fig. \ref{Pseudo_Static}. The radius of the galaxy appears to increase by a factor of 6.8 between $z=3.2$ to $z\approx0$, which is in close agreement with the observed data from the previous studies \citep{Lopez_2010} which state a size evolution of a factor of 6 in the same interval. It is also noted that the size evolution between $z=7.5$ to $z\approx0$ due to pseudo-evolution is by a factor of 23.81. In any case, we do not argue that the universe is static. This model is taken into account to show that pseudo-evolution between the static and $\Lambda$CDM cosmological model does indeed fit the observed data for non-evolving or mildly evolving galaxies.

 \section{Conclusion} \label{Conclusion}
After a short introduction to $R_\text{h}=ct$, Neumann's and $\Lambda$CDM cosmological models, a galaxy with a fixed proper radius of 25 kpc and an absolute magnitude $M_V=-21.5$ is moved in redshift, and its apparent size and apparent magnitude deduced in different cosmological models is noted. This is performed for later comparison of results from individual cosmological models to mixing of different cosmological models leading to pseudo-evolution of a galaxy. Meanwhile, using the apparent measurements from \citet{2018MNRAS.477.3185L} as an example, the proper size has been obtained for the above three cosmological models. Fig. \ref{Proper_size_in_all_models} shows that these data suggest a growth in the proper radius of galaxies by a factor of about 2 from $z\approx8$ to $z\approx1$ and that the here considered cosmological models lead to similar results.

 In the end, a gedankan experiment on using an observers' choice of cosmological model to study the Universe which behaves in reality according to a different cosmology is conducted. In this experiment a galaxy of a fixed radius in the $R_{\text{h}}=ct$ and Neumann's cosmological model is interpreted in the $\Lambda \text{CDM}$ cosmological model. Under such a framework, the galaxy seems to evolve in size by a factor of 1.1 and 1.3 for $R_\text{h}=ct$ and Neumann's cosmological model between $z=7.5$ to $z\approx0$, respectively. For a static Euclidean model, the pseudo-evolution is of a factor of 23.8, which closely matches the observed size evolution with a size evolution by a factor of 6.8 between $z=3.2$ to $z\approx0$ \citep{Lopez_2010}. This evolution is not real, it arises from mixing two different cosmological models. The pseudo-evolution in $R_\text{h}=ct$ and Neumann's cosmological models observed with the $\Lambda \text{CDM}$ model alleviates the problem of galaxy size evolution. However, the pseudo-evolution which appears for a true static Euclidean model observed with a $\Lambda \text{CDM}$ model, completely accounts for the problem of galaxy size evolution as pseudo-evolution. In any case, we do not advocate any cosmological model over another. This contribution is to demonstrate the physical effects of presuming a cosmological model for interpretation, when the universe exhibits a different one. Especially, in the context of galaxy size evolution, the previously not documented effect of pseudo-evolution is emphasized as a possibly important issue to consider when galaxy evolution models do not match observational data. \\

\noindent\textbf{Acknowledgement:} We thank Eric Lerner for his valuable comments.

\bibliography{vasanth_bibtex}

\section{Appendix} \label{section2}
Cosmological perturbation theory is premised upon linear perturbation theory in curved spacetime. The linear perturbation theory consists of a background and a physical spacetime with a perturbation propagating on the background spacetime. The perturbation is seen as the difference between the pulled-back metric of the physical and the background metric, given as
\begin{equation}
{^l}{g}{_{\mu \nu}} = \phi^{*}({g}{_{\mu \nu}})-{^b}{g}{_{\mu \nu}}
\end{equation}
where  ${^l}g{_{\mu \nu}}$ denotes the perturbation, $g_{\mu \nu}$ the physical metric, ${^b}{g}{_{\mu \nu}}$ is the background metric and $\phi^*$ is the pullback of the diffeomorphism $\phi$  . The background spacetime is usually restricted to being a homogeneous and isotropic spacetime as perturbations on this spacetime can be written in terms of harmonic functions, which can be solved readily. Another reason is that on large scales the universe is homogeneous and isotropic.

Also we assume that the universe's departure from homogeneity and isotropy due to the perturbation is very small \citep{Javanmardi_Kroupa_Por,Javanmardi_Kroupa}. This enables us to take the Einstein tensor of the physical metric upto the first order of
the linearized theory. This is valid as we assumed a weak gravitational field implying that the perturbation is small. For the background metric we use the flat FLWR metric with its components given as  
\begin{equation}
\begin{aligned}
 {^b}{g}{_{00}} &=-1, \\
 {^b}{g}{_{ii}} &=R^2(t),
\end{aligned}
\end{equation}
where $R$ is the scale factor as a function of coordinate time $t$, and local perturbations can be decomposed into scalars, vectors, and tensors based on their transformation properties under $3\text{D}$ rotations. Further we take only the scalar perturbation as it takes prominence on cosmological scales over the vector and tensor perturbations since they characterise the energy density of inhomogeneities,
which may play a role in formation of structure in the universe. The vector contribution represents rotation by the cosmological fluid which decays very quickly, thus not contributing on the cosmological scale. The tensor contributions manifest the gravitational waves which does not induce any perturbations on the perfect fluid in linear approximation \citep{Mukhanov}. Therefore the local perturbation metric parametrised with scalar fields $\Phi$ and $\Psi$ under longitudinal (Newtonian) gauge is given as
\begin{equation}
 \begin{aligned}
  {^l}{g}{_{00}} &= -2\Phi,  \\      
  {^l}{g}{_{ii}} &= -2\Psi.
 \end{aligned}    
 \end{equation}
For a derivation of the above equations in this gauge we refer to chapter 7 in \citet{Mukhanov}. For a review on different gauges in cosmological perturbation theory see chapter 4 in \citet{1995STIN...9622249B}, and for a good introduction to cosmological perturbation theory with a more general perturbation we refer to \citet{1980PhRvD..22.1882B}, \citet*{1992PhR...215..203M},  \citet{1984PThPS..78....1K}.

 The Christoffel's symbols of the physical metric upto the first order are given as \citep{Weinberg}  
\begin{equation}
 \begin{aligned}
 {^b}{\Gamma}{^0_{ij}}&=R R^{'} \delta_{ij},\\
 {^b}{\Gamma}{^i_{0j}}&={^b}{\Gamma}{^i_{j0}}=H\delta_{ij},\\
 \end{aligned}
 \end{equation}
and for the local perturbation as \citep{2016Ap&SS.361...93S}
\begin{equation}
 \begin{aligned}
 {^l}{\Gamma}{^i_{00}}&= \frac{\partial_{i} \Phi}{R^2},\\
 {^l}{\Gamma}{^0_{i0}}&= \partial_{i}\Phi,\\
 {^l}{\Gamma}{^0_{ij}}&= {-\delta}{_{ij}(2R R^{'}\Phi+\Psi^{'})},\\
 {^l}{\Gamma}{^i_{j0}}&= \frac{\delta_{ij}}{R^2}(2H\Psi- \Psi^{'}),\\
 {^l}{\Gamma}{^i_{jk}}&= \frac{1}{R^2}(\delta_{jk} \partial_i \Psi-\delta_{ij}\partial_k \Psi- \delta_{ik}\partial_{j}\Psi).\\
 \end{aligned}
 \end{equation}

 For the sake of brevity we will not explicitly write the Ricci tensor and Riemann tensor. For a detailed review on them refer to \citet{1980PhRvD..22.1882B} and \citet{2016Ap&SS.361...93S}. The components of the background Einstein's tensor \citep{Mukhanov} are
 \begin{equation}
 \begin{aligned}
         {^b}{G}_{00}&=3H^2,\\
         {^b}{G}_{ij}&=(2H^{'}+H^{2})\delta_{ij},
 \label{backgroundEinsteintensor}
 \end{aligned}
 \end{equation}
 and the components of the local Einstein's tensors are \citep{2016Ap&SS.361...93S}  
 \begin{equation}
 \begin{aligned}
{^l}{G}_{00}&= \frac{2}{R^4}(6 \Phi R^{'2} +\bigtriangledown^2 \Psi - 3RR^{'} \Psi^{'}), \\
 {^l}{G}_{0j}&= \frac{R^{'}}{R} \partial_{j}\Phi - \frac{R^{'}}{R^{3}}4 \partial_{j} \Psi + \frac{2}{R^2} \partial_j \Psi^{'}, \\
\label{l^G_00}
 \end{aligned}
 \end{equation}
 \begin{equation}
 \begin{aligned}
 \label{offdiagonal of G}
 {^l}{G}_{ij}&= (\delta_{ij} \bigtriangledown^{2} -\partial_{i} \partial_{j}) \Bigg( \Phi- \frac{\Psi}{R^2} \Bigg)+ 2\delta_{ij} \Psi^{'2}+ 2 \delta_{ij} R R^{'} \Phi^{'}\\ & + 4 \delta_{ij} R R^{''} \Phi + 2\delta_{ij} R^{'2} \Phi -2\delta_{ij} \frac{R^{'}}{R} \Bigg( \Psi^{'}- \frac{R^{'}}{R} \Psi \Bigg). 
 \end{aligned}
 \end{equation}
 The primes on the top of the symbols represent coordinate time derivatives. The rotational and translational invariance of the unperturbed energy\texttt{-}momentum tensor requires it to be a perfect fluid \citep{Weinberg}, and for the perturbed metric we assume a general energy\texttt{-}momentum tensor. To establish a relationship between the two scalar fields we take off-diagonal elements of Eq. \ref{offdiagonal of G} and set the anisotropic inertia terms of the perturbed energy\texttt{--}momentum tensor to zero,  
\begin{equation}
\partial_{i} \partial_{j} \Bigg(\Phi-\frac{\Psi}{R^2} \Bigg)=0.
\label{laplace}
\end{equation}
This is the famous result of cosmological perturbation theory in longitudinal (Newtonian) gauge. We refer the reader to \citet{Weinberg} eq. 5.3.20, and \citet{Mukhanov} eq. 7.46 for a detailed derivation of Eq.\ref{laplace} in longitudinal (Newtonian) gauge. We obtain a similar result for a static perturbation theory in Minkowski background metric with transverse gauge (generalisation of the longitudinal gauge) in  eq. 7.56 of \cite{SeanCarroll}. 

Now to solve Eq. \ref{laplace}, we first fix the boundary conditions. Here we have four different boundary conditions: the Dirichlet B.C. where the function is fixed at the boundary, the Neumann B.C. where the normal derivative of the functions is fixed at the boundary, the Cauchy B.C. where the the function and its normal derivative is fixed at the boundary, and lastly the Mixed B.C, e.g. a Dirichlet B.C. on the northern hemisphere and Neumann B.C. in the southern hemisphere.

For our case we will use the Neumann B.C. since the gravitational force is zero at the boundary \citep*{2016Ap&SS.361...93S} and the Cauchy B.C. removes interesting dynamics from the system that we would like to study \citep*{2017MPLA...3250077K} while the Mixed B.C. does not agree with our fundamental assumption of spacetime isotropy. The Dirichlet B.C. is the usually preferred solution in the literature, however the Neumann B.C. is equally stable on a closed surface, see chapter 9 in \citet{Arfken}. Therefore solving Eq.\ref{laplace} using the Neumann B.C. we obtain the solution as
\begin{equation}
 \begin{aligned}
 \Phi=\frac{\Psi}{R^2}+c_1.  
\label{solution_to_laplace}
 \end{aligned}
 \end{equation}
The general solution to Eq. \ref{laplace} using the Neumann B.C. is
\begin{equation}
 \begin{aligned}
 \Phi-\frac{\Psi}{R^2}=A_1(t,x)+A_2(t,y)+A_3(t,z)+c_1,
 \end{aligned}
 \end{equation}
where the homogeneity and isotropy of the perturbations set the first three terms to zero yielding Eq. \ref{solution_to_laplace}. The constant $c_1$ is the spatial average of the perturbations, setting it to zero is equivalent to considering the Dirichlet B.C. \citep{1992PhR...215..203M}. Through out this paper we will use the Neumann B.C. with the solution given by Eq. \ref{solution_to_laplace}.
 The interpretation for the constant $c_1$ is still an open question. For an experimental determination of the value of $c_1$ using strong lensing systems see \citet{2016Ap&SS.361...93S}.

\bsp	
\label{lastpage}
\end{document}